\documentclass[12pt,a4paper,twocolumn]{article}
\usepackage{amsmath,amssymb,theorem}
\begin{document}
%\layoutstyle{aipproc.sty}
\date{}
\title{On the Masses of the Universal hypermultiplets in heterotic M-theory}
\author{Nasr Ahmed\\ \small{Astronomy Department, National Research Institute of Astronomy And Geophysics,}\\ \small{Helwan, Cairo, Egypt.}}
\maketitle
\section{abstract}
The reduced 5D Heterotic M-theory has a deeply rich structure. For every Calabi-yau compactification, there exists a gravitational hypermultiplet $(g_{\mu\nu},\psi_{\mu},A_{\mu})$ and a universal hypermultiplet. In this paper we derive the formulae for the masses of the scalar sector of the universal hypermultiplet $(V,\sigma,\zeta,\bar{\zeta})$ in the framework of 5D Heterotic M-theory. 
\section{Introduction}
 
In the original formulation of M-theory \cite{(5),(5a)}, all the standard model fields are trapped on two 
9-branes located at the end points of an $S^{1}/Z_{2}$ orbifold. The 6 extra dimensions on the branes are compactified. A 5 dimensional reduction of the original Horava–Witten theory and the corresponding braneworld
cosmology is given in \cite{(6),(7),(8)}. 

In the 11-dimensional theory, the supergravity multiplet consists of the graviton field or the metric $g$, gravitino field $\psi_I$ and a three index gauge field $C_{IJK}$ with a field strength $G_{IJKL}$. 
The total bulk field content of this 5 dimensional theory is given by the gravity multiplet $(g_{\alpha \beta}, A_{\alpha}, \psi^{i}_{\alpha})$ together with the universal hypermultiplet $(V,\sigma, \zeta, \bar{\zeta})$. $A_{\alpha}$ is a gauge field with field strength $\bar{F}_{\alpha \beta}=\partial_{\alpha} A_{\beta}-\partial_{\beta} A_{\alpha}$, $\zeta$ is a background complex field, $V$ is the Calabi-Yau volume and $\psi^{i}_{\alpha}$ is the gravitino field.
After the dualization, the three-form $C_{\alpha \beta \gamma}$ produces a scalar field $\sigma$. The 5 dimensional effective action can be written as \cite{(7)}
\vspace{1.0 mm}
\begin{equation}
S_{5}=S_{bulk}+S_{bound}
\end{equation}
where
\begin{eqnarray}
S_{bulk}=\frac{-1}{2\kappa_{5}^{2}}\int_{{\cal M}_{5}}\sqrt{-g} \left[{\cal R}+\frac{3}{2}\bar{F}_{\alpha \beta}\bar{F}^{\alpha \beta} \right. \\ \nonumber
\left. +\frac{1}{\sqrt{2}}\epsilon^{\alpha \beta \gamma \delta \epsilon} A_{\alpha}\bar{F}_{\beta \gamma}\bar{F}_{\delta 
\epsilon}+ \frac{1}{2V^{2}}\partial_{\alpha}V \partial^{\alpha}V \right. \\ \nonumber
 \left. +\frac{1}{2V^{2}}\left[(\partial_{\alpha}\sigma -i(\zeta \partial_{\alpha}\bar{\zeta}-\bar{\zeta}\partial_{\alpha}\zeta) \right.\right. \\ \nonumber
 \left.\left. -2\alpha \epsilon (x^{11}) A_{\alpha})\right]^2 +\frac{2}{V} \partial_{\alpha}\zeta \partial^{\alpha}\bar{\zeta}+\frac{\alpha^{2}}{3V^{2}} \right]
\end{eqnarray} 
and
\begin{eqnarray} \label{boundddd} 
S_{bound}=\frac{\sqrt{2}}{\kappa_{5}^{2}}\left[\int_{M_{4}^{(1)}}\sqrt{-g}V^{-1}\alpha - \right.\\ \nonumber
\left. \int_{M_{4}^{(2)}}\sqrt{-g}V^{-1}\alpha \right] -\frac{1}{16\pi \alpha_{GUT}} \\  \nonumber
 \sum_{i=1}^{2}\int_{M_{4}^{(i)}}\sqrt{-g}\left(V \text{tr}F_{\mu\nu}^{(i)}F^{(i)\mu\nu} \right. \\  \nonumber
\left. - \sigma \text{tr}F_{\mu\nu}^{(i)}\widetilde{F}^{(i)\mu\nu}\right). 
\end{eqnarray}
where $\widetilde{F}^{(i)\mu\nu}=\frac{1}{2}\epsilon^{\mu \nu \rho \sigma} F_{\rho \sigma}^{(i)}$ and the expansion coeffecients $\alpha_i$ are
\begin{eqnarray}
\alpha_i=\frac{\pi}{\sqrt{2}}\left(\frac{\kappa}{4\pi}\right)^{2/3}\frac{1}{v^{2/3}}\beta_i,\\  \nonumber
\quad \beta_i=-\frac{1}{8\pi^2}\int_{C_i}tr({\cal R}\wedge{\cal R}).
\end{eqnarray}
with the Calabi-Yau volume $V$ defined as
\begin{equation}
V=\frac{1}{v}\int_{X}\sqrt{g^{(6)}}
\end{equation}
where $g^{(6)}$ is the determinant of the Calabi-Yau metric.

\section{The Non-linear Sigma model Lagrangian for the background $V$ and $\zeta$ fields} \label{contrib}
In gaugino condensates the gaugino acquires non-zero vacuum expectation value which breaks the supersymmetry. The gaugino condensates lead to a background $\zeta$ field. So, now we have a background $V$ field (represents the size of Calabi-Yau space), a background $\zeta$ field and a background metric. We take the following nonlinear sigma model lagrangian 
\begin{eqnarray}
L=\frac{-(\partial V)^{2}-\left(\partial \sigma-i(\zeta \partial \bar{\zeta}-\bar{\zeta}\partial \zeta)\right)^{2}}{4V^{2}} \\  \nonumber 
-\frac{1}{V}(\partial \zeta)(\partial \bar{\zeta})-U  \label{nonlinear}~~~~~~~~~~~~~~~~~~~~
\end{eqnarray}
Where $U=\frac{\alpha^{2}}{6V^{2}}$.
We define the metric as 
\begin{equation}
dh^{2}=\frac{dV^{2}}{4V^{2}}+\frac{\left(d\sigma-i(\zeta d\bar{\zeta}-\bar{\zeta} d\zeta)\right)^{2}}{4V^{2}}+\frac{d\zeta d\bar{\zeta}}{V}
\end{equation}
and introduce the one-forms $u$ and $v$ with their complex conjugate $\bar{u}$ and $\bar{v}$
\begin{equation}
w^{a}=(u,\bar{u},v,\bar{v})
\end{equation}
Where
\begin{equation}
u=\frac{d\zeta}{\sqrt{V}}
\end{equation}
And
\begin{equation}
v=\frac{1}{2V}\left(dV+id\sigma+\zeta d\bar{\zeta}-\bar{\zeta} d\zeta\right)
\end{equation}
This leads to
\begin{equation}
du=-\frac{1}{2}(v+\bar{v})\wedge u
\end{equation}
\begin{equation}
dv=-\bar{v}\wedge v-\bar{u}\wedge u
\end{equation}
and the connection two-forms
\begin{eqnarray}
w^{v}~_{v}&=&\bar{v}-v \\
w^{u}~_{u}&=&\frac{1}{2}(\bar{v}-v)\\
w^{u}~_{v}&=&-u
\end{eqnarray}
Cartan's structure equantions are
\begin{eqnarray}
T^{a}&=&d\theta^{a}+w^{a}_{b}\wedge \theta^{b}\\
\Omega^{a}_{b}&=&dw^{a}_{b}+w^{a}_{c}\wedge w^{c}_{b}
\end{eqnarray}
For the torsion and curvature 2-form respectively.\\
The curvature 2-form gives (when expressed locally)
\begin{equation}
\Omega^{a}_{b}=\frac{1}{2}R^{a}_{bcd} \theta^{c} \wedge \theta^{d}
\end{equation}
Where the components $R^{a}_{bcd}$ are in orthonormal basis.
The affine connection form satisfies
\begin{equation}
w_{ab}+w_{ba}=dg_{ab}~~,~~w_{ab}=g_{ac}w^{c}_{b}
\end{equation}
Since for the orthonormal metric $g_{ab}=\eta_{ab}$ is constant, we have
For the curvature 2-form we have
\begin{equation}
\Omega_{ab}=g_{ac}\Omega^{c}_{b}=-\Omega_{ba}
\end{equation}
For a free torsion space, Cartan's first structure equation is
\begin{equation}
d\theta^{a}=-w^{a}_{b}\wedge \theta^{b}
\end{equation}
So, the key formulas we are going to use to derive the connection coefficients and the corresponding curvature tensor are
\begin{eqnarray}
w_{ab}&=&-w_{ba} \\ 
d\theta^{a}&=&-w^{a}_{b}\wedge \theta^{b} \\
\frac{1}{2}R^{a}_{bcd} \theta^{c} \wedge \theta^{d}&=&dw^{a}_{b}+w^{a}_{c}\wedge w^{c}_{b}
\end{eqnarray}
For the Ricci tensor we find
\begin{eqnarray}
R^{u}~_{u}=\frac{1}{2}R^{u}~_{u \alpha \beta} w^{\alpha} \wedge w^{\beta}\\ \nonumber
=(\bar{v} \wedge v)+(\bar{u} \wedge u)
\end{eqnarray}
And
\begin{eqnarray}
R^{v}~_{v}=(\bar{u} \wedge u)-(\bar{v} \wedge v)
\end{eqnarray}
The  components of the curvature tensor are ($g_{u\bar{u}}=\frac{1}{2}$):
\begin{eqnarray} \nonumber
R_{u\bar{u}u\bar{u}}=1,~~R_{u\bar{u}v\bar{v}}=\frac{1}{2},~~R_{v\bar{v}v\bar{v}}=1, ~~~~~~~~~\\  \nonumber
R_{u\bar{v}v\bar{u}}=\frac{1}{2},~~R_{\bar{u}uu\bar{u}}=-1,~~R_{u\bar{u}\bar{u}u}=-1,~~~~~~~~~\\  \nonumber
R_{\bar{u}uv\bar{v}}=-\frac{1}{2},~R_{u\bar{u}\bar{v}v}=-\frac{1}{2},~~R_{\bar{v}uv\bar{u}}=-\frac{1}{2},~~~~~~~~\\ \nonumber
~~R_{\bar{v}vv\bar{v}}=-1,~~R_{u\bar{v}\bar{u}v}=-\frac{1}{2},~~R_{v\bar{u}u\bar{v}}=\frac{1}{2},~~~~~~~~\\  \nonumber
~~R_{\bar{u}u\bar{u}u}=1,~~R_{\bar{v}v\bar{v}v}=1,~~R_{\bar{u}u\bar{v}v}=\frac{1}{2},~~~~~~~~\\ \nonumber
~~R_{\bar{v}vu\bar{u}}=-\frac{1}{2},~~R_{v\bar{v}\bar{u}u}=-\frac{1}{2},~~R_{\bar{v}v\bar{u}u}=\frac{1}{2},~~~~~~~~\\ \nonumber
~~R_{\bar{v}u\bar{u}v}=\frac{1}{2},~~R_{\bar{u}vu\bar{v}}=\frac{1}{2},~~R_{v\bar{u}\bar{v}u}=-\frac{1}{2},~~~~~~~~\\ \nonumber
~~R_{\bar{u}v\bar{v}u}=\frac{1}{2},  ~~R_{v\bar{v}u\bar{u}}=\frac{1}{2}R_{v\bar{v}\bar{v}v}=-1.~~~~~~~~
\end{eqnarray}
The grad of the potential $U$ is
\begin{equation}
\nabla U=-\frac{\alpha^{2}}{3V^{2}}(v+\bar{v})
\end{equation}
\begin{eqnarray}
\nabla^{2} U=\frac{2\alpha^{2}}{3V^{2}}(v+\bar{v})\otimes (v+\bar{v})-\frac{\alpha^{2}}{3V^{2}} \\ \nonumber
(v-\bar{v})\otimes(v-\bar{v})+\frac{2\alpha^{2}}{3V^{2}}u\otimes \bar{u}
\end{eqnarray}
Now we would like to express the the lagrangian (\ref{nonlinear}) in terms of the one-forms $u$ and $v$. We make use of the general form of the nonlinear sigma model lagrangian with a background field $\zeta$ 
\begin{eqnarray}
L&=&-\frac{1}{2}g_{ij}(D_{\mu}\zeta^{i})(D^{\mu}\zeta^{j})\\ \nonumber
&&+\frac{1}{2}(\partial_{\mu}\phi^{i})(\partial^{\mu}\phi^{j})R_{ikjl}\zeta^{k}\zeta^{l}-\frac{1}{2}U_{;ij}\zeta^{i}\zeta^{j}.
\end{eqnarray}
Where $R_{ijkl}$ is the curvature of $g_{ij}$. After some manipulations we get the lagrangian in the form
\begin{equation}
L=L_{1}+L_{2}
\end{equation}
Where
\begin{equation}
L_{1}=-(D_{\alpha}\zeta^{u})(D^{\alpha}\zeta^{\bar{u}})-(D_{\alpha}\zeta^{v})(D^{\alpha}\zeta^{\bar{v}})~~~~~~~~~~~~~~~~~~~~~~~~~~~~~~~~~~~~~\\
\end{equation}
And
\begin{eqnarray}
L_{2}=\frac{\alpha^{2}}{2}V^{-2}\left[(\zeta^{v}-\zeta^{\bar{v}})+ C(\zeta^{u}-\zeta^{\bar{u}})\right)]^{2} \\ \nonumber
-\frac{\alpha^{2}}{2}V^{-2}(\zeta^{u}-C\zeta^{v})(\zeta^{\bar{u}}-C\zeta^{\bar{v}})~~~~~~\\ \nonumber
-\frac{\alpha^{2}}{2}V^{-2}(\zeta^{v}+\zeta^{\bar{v}})^{2}-\frac{\alpha^{2}}{3}V^{-2}\zeta^{u}\zeta^{\bar{u}}~~~
\end{eqnarray}
The first part $L_{1}$ is diagonalized and we need to diagonalize the second part $L_{2}$ to get the eigen Values. To do that we make the following change of variables
\begin{equation}
\zeta^{v}=\frac{1}{\sqrt{2}}(X+iY)~~,~~\zeta^{u}=\frac{1}{\sqrt{2}}(Z+iW)
\end{equation}
That means we have 4 fields X,Y,Z, and W. In terms of the new fields, the lagrangian $L_{2}$ becomes 
\begin{eqnarray}
L_{2}&=&-\alpha^{2}V^{-2}(Y+CW)^{2} \\ \nonumber
&-&\frac{\alpha^{2}}{4}V^{-2}\left((Z-CX)^{2}+(W-CY)^{2}\right)\\ \nonumber
&-&\frac{\alpha^{2}}{3}V^{-2}\left(2X^{2}+Y^{2}+\frac{1}{2}Z^{2}+\frac{1}{2}W^{2}\right).
\end{eqnarray}
Which could be written in a matrix form as
\[ L_{2}=-\frac{\alpha^{2}}{V^{2}}\left( \begin{array}{cccc}
\frac{2}{3}+\frac{C^{2}}{4} & 0 & -\frac{C}{2}& 0   \\
0 & \frac{4}{3}+\frac{C^{2}}{4}&0& 2C   \\
-\frac{C}{2}&0&\frac{5}{12}&0    \\ 
0& -\frac{C}{2}&0& \frac{5}{12}+ C^{2}\end{array} \right)\] 
After diagonalization , The eigen values that represents the masses of the scalar sector of the universal hypermultiplet $(V,\sigma,\zeta,\bar{\zeta})$ are 
\begin{eqnarray}
\frac{\alpha^{2}}{24V^{2}}\left(-21-15C^{2}+\sqrt{121-774C^{2}+81C^{4}}\right),\\
\frac{\alpha^{2}}{24V^{2}}\left(-21-15C^{2}+\sqrt{121-774C^{2}+81C^{4}}\right),\\
\frac{\alpha^{2}}{24V^{2}}\left(-13-3C^{2}+3\sqrt{1+18C^{2}+C^{4}}\right),\\
\frac{\alpha^{2}}{24V^{2}}\left(-13-3C^{2}-3\sqrt{1+18C^{2}+C^{4}}\right).
\end{eqnarray}

\section{Conclusion}
Making use of the non-linear sigma model Lagrangian of the background fields, We derived the formulae for the masses of the scalar sector of the universal hypermultiplet $(V,\sigma,\zeta,\bar{\zeta})$ in the reduced 5D Heterotic M-theory.
\section{Acknwoledgment}
I am so grateful to Prof. Ian Moss from Newcastle university for the very useful discussions during this work.

\end{document}